\documentclass[]{aa} % for the letters 
\usepackage{graphicx}
\usepackage{txfonts}
\usepackage[breaklinks=true]{hyperref} %% to avoid \citeads line fills
\usepackage{natbib,twoopt}
\usepackage{float}
\usepackage{arydshln}
\usepackage[breaklinks=true]{hyperref} %% to avoid \citeads line fills
\bibpunct{(}{)}{;}{a}{}{,}    %% natbib format like A&A and ApJ
\newcommand{\MJup}{\ensuremath{M_{\mathrm{Jup}}}\xspace}

\newcommand{\MSun}{\ensuremath{M_{\odot}}\xspace}

\newcommand{\LSun}{\ensuremath{L_{\odot}}\xspace}
\newcommand{\mic}{$\mu$m\xspace}
\begin{document} 

\title{Constraining the mass of the planet(s) sculpting a disk cavity.
    \thanks{Based on observations made with the \textit{VLT}, program 095.C-0673(A).}
}
\subtitle{The intriguing case of 2MASS J16042165-2130284}

\author{H. Canovas\inst{1,2,9}, 
   A. Hardy\inst{2,9},
   A. Zurlo\inst{3,9},
   Z. Wahhaj\inst{4},
   M.~R. Schreiber\inst{2,9},
   A. Vigan\inst{5,4},
   E. Villaver\inst{1},
   J. Olofsson\inst{2,9},
   G. Meeus\inst{1},
   F. M\'enard\inst{6,7},
   C. Caceres\inst{2,8,9},
   L.~A. Cieza\inst{3,9},
   A. Garufi\inst{1}
}

\institute{ 
Departamento de F\'isica Te\'orica, Universidad Aut\'onoma de Madrid, Cantoblanco, 28049, Madrid, Spain.\\
\email{hector.canovas@uam.es} 
\and
Departamento de F\'isica y Astronom\'ia, Universidad de Valpara\'iso, Valpara\'iso, Chile.
\and
Facultad de Ingenier\'ia, Universidad Diego Portales, Av. Ejercito 441, Santiago, Chile.
\and
European Southern Observatory, Alonso de Cordova 3107, Vitacura, Santiago, Chile.
\and
Aix Marseille Universit\'e, CNRS, LAM (Laboratoire d'Astrophysique de Marseille) UMR 7326, 13388, Marseille, France.
\and
CNRS, IPAG, F-38000 Grenoble, France.
\and
Univ. Grenoble Alpes, IPAG, F-38000 Grenoble, France.
\and
Departamento de Ciencias Fisicas, Facultad de Ciencias Exactas, Universidad Andres Bello. Av. Fernandez Concha 700, Las Condes, Santiago, Chile.
\and
Millennium Nucleus ``Protoplanetary Disks in ALMA Early Science".
}

\date{\today}

\abstract
% context heading (optional)
{The large cavities observed in the dust and gas distributions of transition disks may be explained by planet-disk interactions. At $\sim145$ pc,
2MASS\,J16042165-2130284 (J1604) is a 5-12 Myr old transitional disk with different gap sizes in the mm- and $\mu$m-sized dust distributions
(outer edges at $\sim79$ and at $\sim63$ au, respectively). Its $^{12}$CO emission shows a $\sim30$ au cavity. This radial structure suggests
that giant planets are sculpting this disk.}
% aims heading (mandatory)
{We aim to constrain the masses and locations of plausible giant planets around J1604.}
% methods heading (mandatory)
{We observed J1604 with the Spectro-Polarimetric High-contrast Exoplanet REsearch (SPHERE) at the Very Large Telescope (VLT),
in \texttt{IRDIFS\_EXT}, pupil-stabilized mode, obtaining $YJH$- band images with the integral field spectrograph (IFS) and $K1K2$-band
images with the Infra-Red Dual-beam Imager and Spectrograph (IRDIS). The dataset was processed exploiting the angular differential imaging
(ADI) technique with high-contrast algorithms.}
% results heading (mandatory)
{Our observations reach a contrast of $\Delta K, \Delta YH\sim12$ mag from $0\farcs15$ to $0\farcs80$ ($\sim22$ to 115 au), but no planet
candidate is detected. The disk is directly imaged in scattered light at all bands from $Y$ to $K$, and it shows a red color. This indicates that
the dust particles in the disk surface are mainly $\gtrsim0.3\,\mu$m-sized grains. We confirm the sharp dip/decrement in scattered light in
agreement with polarized light observations. Comparing our images with a radiative transfer model we argue that the southern side of the disk
is most likely the nearest.}
% conclusions heading (optional), leave it empty if necessary 
{This work represents the deepest search yet for companions around J1604. We reach a mass sensitivity of $\gtrsim 2-3\MJup$ from $\sim22$
to $\sim115$ au according to a hot start scenario. We
propose that a brown dwarf orbiting inside of $\sim 15$ au and additional Jovian planets at larger radii could account for the observed
properties of J1604 while explaining our lack of detection.}


\titlerunning{J1604}
\authorrunning{Canovas et al.}

\maketitle

%__________________________________________________________________________________________
\section{Introduction}
\label{sec:intro}

While there is virtually no doubt that giant planets (here we follow the broad definition of planet adopted by
the Working Group on Extrasolar Planets of the International Astronomical Union \citep{2007IAUTA..26..183B})
form in protoplanetary disks, when, where, and under which conditions this occurs, is still highly uncertain. A greater understanding of this
process is needed in order to explain the properties of the formed planets and the diverse architecture of planetary
systems including that of our Solar System \citep[e.g., ][and references therein]{2015ARA&amp;A..53..409W}.
Early studies of planet disk interactions predict that forming giant planets open gaps and cavities in the surrounding
gaseous disks \citep[e.g., ][]{1994ApJ...421..651A}, while recent detailed models predict a variety of additional effects related to
planet-disk interactions such as gas flows/accretion streams through the disk cavity \citep{1999ApJ...526.1001L}, dust filtration 
\citep{2006A&amp;A...453.1129P, 2006MNRAS.373.1619R}, a decrease in the accretion rate of gas onto the star 
\citep{2007MNRAS.375..500A}, an accumulation of millimeter sized particles in a narrow ring at the outer edge of the cavity
\citep{2012A&amp;A...545A..81P}, and/or spiral arms detectable in scattered light \citep{2015MNRAS.451.1147J, 2015ApJ...809L...5D}.
Many of these features have been observed \citep[e.g.,][]{2011ApJ...732...42A, 2015ApJ...805...21C, 2016MNRAS.458L..29C, 2013Natur.493..191C, 
2013A&amp;A...560A.105G, 2016ApJ...816L..12D, 2009ApJ...698..131H, 2016A&amp;A...585A..58V, 2015A&amp;A...584L...4P},
however despite numerous observational efforts, definitive detections of planets at their early formation stage have proven to be
extremely hard to achieve. Potential detections based on aperture masking techniques \citep[e.g., ][]{2011A&A...528L...7H} may
actually be tracing scattered light from the inner disk regions \citep[][]{2012ApJ...752...75C, 2013A&amp;A...552A...4O}. To date only
two disks show plausible detections of embedded planet companions: LkCa 15 \citep{2012ApJ...745....5K, 2015Natur.527..342S},
and HD\,100546 \citep{2013ApJ...766L...1Q, 2015ApJ...807...64Q, 2014ApJ...796L..30C, 2015ApJ...814L..27C}, while the disk HD\,169142
shows evidences of an embedded brown dwarf \citep{2014ApJ...792L..22B, 2014ApJ...792L..23R}. Probably the least ambiguous detections
of young planets are those located at large separations (>100 au) from their host stars \citep[e.g., ][]{2014ApJ...781...20K, 2015ApJ...806L..22C}.
Their formation via core-accretion \citep{1996Icar..124...62P} is unlikely \citep[e.g.,][]{2011ApJ...727...86R} but in principle they could form
via gravitational instability \citep{1997Sci...276.1836B}. As long as we lack a direct detection of a forming planet inside a disk gap/cavity,
observed features such as dust filtering and piling up of large grains in ring-like structures cannot be conclusively attributed  to the presence
of planets. In order to confront models of planet-disk interactions, either planet detections or deep detection limits on the presence of
planets inside the disk cavity must be obtained. One disk which presents an opportunity to conduct such a study is 2MASS J16042165-2130284
(hereafter referred to as J1604).

J1604 ($m_\mathrm{R, H, K} = 11.8, 9.1, 8.5$) is a pre-main sequence star which belongs to the Upper Scorpius subgroup (USco) of the
Scorpius-Centaurus OB association at an average distance of $145\pm2$ pc \citep{1999AJ....117..354D}. Age estimations for this association range
from 5 to 12 Myr \citep{1989A&A...216...44D, 2002AJ....124..404P, 2012ApJ...746..154P}. \citet{2014ApJ...787...42C} describes J1604 as a  K2 star
with temperature $T_\mathrm{eff} = 4898^{+184}_{-177}\,K$, luminosity $L_{\star} = 0.58^{+0.22}_{-0.16}$ $\LSun$, and mass
$M_{\star} = 0.95^{+0.12}_{-0.08}$ $\MSun$. The spectral energy distribution (SED) of J1604 shows a deficit of flux excess at near-infrared (NIR)
wavelengths characteristic of transition disks \citep{1989AJ.....97.1451S}, followed by a strong rising excess at 16 $\mu$m and beyond
\citep{2006ApJ...651L..49C, 2009ApJ...705.1646C, 2009AJ....137.4024D} indicating emission from an inner wall directly exposed to stellar radiation at
large separations ($\sim$ tens of au) from the star. Its small (yet significant) excess between $3$ and $16$\,$\mu$m traces optically thick dust at $\sim0.1$ au
\citep[$T\sim900$ K, ][]{2013A&A...558A..66M, 2015A&amp;A...579A.106V}. Therefore J1604 is a `pre-transitional disk', that is, it has optically thick inner
and outer disks separated by an optically thin gap \citep{2007ApJ...670L.135E}. J1604 is surrounded by the most massive disk known in the USco region
\citep[$\sim0.1\,\MJup$ in dust, ][]{2012ApJ...753...59M, 2014ApJ...787...42C} and this disk has been resolved in scattered polarized light at optical
\citep{2015A&amp;A...584L...4P} and NIR wavelengths \citep{2012ApJ...760L..26M}, and in thermal emission at sub-mm wavelengths
\citep{2012ApJ...745...23M, 2014ApJ...791...42Z}. These observations reveal an almost face-on disk \citep[$i\sim6\degr$,][]{2012ApJ...753...59M}
at position angle  $PA\sim77\degr$ \citep[east of north,][]{2014ApJ...791...42Z} with a large gap. Interestingly, the inferred gap outer edge is
wavelength-dependent: $\mathrm{\sim79}$ au at sub-mm wavelengths and $\mathrm{\sim63}$ au in the NIR, which may be evidence of the dust filtration
produced by planets, as predicted by \citet{2006MNRAS.373.1619R}. The polarized images of J1604 also reveal an intriguing, sharp decrement in emission
along the bright inner rim \citep{2012ApJ...760L..26M, 2015A&amp;A...584L...4P}. ALMA $^{12} \mathrm{CO}\,(2-1)$ observations reveal a large gaseous
disk with a $\sim30$ au cavity in radius \citep[][]{2014ApJ...791...42Z, 2015A&amp;A...579A.106V}. 

As the disk morphology of J1604 shows several structures generally attributed to planet-disk interactions, it has been the subject of many searches for
low-mass companions in the past. RX\,J1604.3-2130B (M2, $M_\star \sim0.7\,\MSun$) is the only known companion around J1604, located at a projected
separation of 2350 au \citep[][]{2009ApJ...703.1511K}. Combining aperture masking interferometry with direct imaging \cite{2008ApJ...679..762K} reached
a $\Delta K$ contrast of $\sim3.6$ mag at 10--20 mas (1.5--2.9 au),  $\sim5.4$ mag at 20--40 mas (2.9--5.8 au), and $\sim6.2$ mag at 40--160 mas (5.8--23.2 au).
Assuming a hot start scenario and using 5 Myr isochrones these values were converted to mass limits of  $\sim 70 \MJup$, $\sim 21 \MJup$, and $\sim 15 \MJup$,
respectively. In a complementary survey \citet{2011ApJ...726..113I} reached a $\Delta K$ contrast ranging from $3.4$  to  $5.8$ mag at $300 - 1000$ mas
($45 - 150$ au), yielding mass sensitivities of $83 - 19$ $\MJup$.

In this paper, we present high-contrast observations at NIR bands from $Y$ to $K$ of the transition disk around J1604 obtained with the Spectro-Polarimetric
High-contrast Exoplanet REsearch \citep[SPHERE,][]{2008SPIE.7014E..18B} instrument at the Very Large Telescope (VLT). The remarkable performance of
SPHERE allows us to reach a new contrast domain. No giant planets are detected in our observations, yet we derive strict constraints on the masses of planets that might orbit the star at
20-130 au, reaching a mass sensitivity of $\gtrsim2-3 \MJup$ according to the BT-Settl models \citep{2011ASPC..448...91A}. The high quality of our images
allow us to detect the disk in scattered light at all wavelengths from $Y$ to $K$ bands. Therefore, we also analyse the radial and azimuthal structure of this
interesting disk.

%__________________________________________________________________________________________
\section{Observations}
\label{sec:obs}

J1604 was observed with SPHERE under ESO programme 095.C-0673 (P.I. A. Hardy) during June
10, 2015, and August 13, 2015. The June observations had better weather conditions and reached deeper contrast than the August
observations, therefore we focus on the description and analysis of the first dataset. The instrument was used in the \texttt{IRDIFS\_EXT}
mode, which allows simultaneous observation with the integral-field spectrometer IFS \citep{2008SPIE.7014E..3EC} and with the
differential imager and spectrograph IRDIS \citep{2008SPIE.7014E..3LD}. In this mode the IFS covers a wavelength range from
0.96 to 1.68 \mic ($Y$--$H$~bands), having a spectral resolution of $R\sim$30 inside a $1.7\arcsec\times1.7\arcsec$ field of view
(FoV). The IRDIS sub-system, on the other hand, was configured in its dual-band imaging mode \citep[DBI, ][]{2010MNRAS.407...71V}
to simultaneously observe in $K1$ and $K2$ filters ($\lambda_{K1} = 2.110$~\mic, $\lambda_{K2} = 2.251$~\mic, width $\sim$0.1~\mic)
with a $11.0\arcsec\times11.0\arcsec$ FoV.

Our main goal was to detect young planets inside the disk cavity. To that end, the observations  were carried out in pupil-stabilized
mode to perform angular differential imaging \citep[ADI;][]{2006ApJ...641..556M}. We used an apodized pupil Lyot coronagraph 
\citep{2005ApJ...618L.161S, 2011ExA....30...39C} in combination with a focal plane mask of diameter $0\farcs185$ that provides an
inner working angle (IWA) of $\mathrm{\sim0\farcs092}$ (i.e., 13 au at 145 pc). The complete observation sequence amounted to 2
hours in total (1.6 hours on-source) at airmass ranging from 1.001 to 1.071, resulting in $145\degr$ of sky rotation. We used
detector integration times (DITs) of 32~s for both the IRDIS and IFS instruments.
A $2\times2$ dithering pattern was applied to average out flat-field imperfections of the IRDIS detector. The weather conditions were good
with median seeing of $1.0\arcsec$ and coherence time of $\tau_{0} = 2.0$ ms. The wind speed was mostly constant with a median
value of 10.9 m $\mathrm{s^{-1}}$. 

A total of 40 unsaturated (0.8s integration time), off-mask stellar point-spread functions (PSFs) of J1604 were acquired to allow for
relative flux calibration. Sky background images were recorded at the beginning and the end of the sequence. Centering frames with
four centro-symmetric satellite spots (artificial replicas of the central star at different positions on the detector, used to accurately
find the position of the star behind the coronagraph mask) centered on the star
were interleaved with the science frames to accurately derive the position of the star in the detector behind the focal plane. Standard
calibrations were obtained the following morning as part of the general SPHERE calibration strategy.

%__________________________________________________________________________________________
\section{Data reduction}
\label{sec:data_reduction}
In this section we briefly describe the steps followed to clean and align the raw data before applying dedicated algorithms to post-process 
the data. We used the SPHERE data reduction and handling (DRH) pipeline \citep[v. 0.15.0;][]{2008SPIE.7019E..39P}, as well as additional,
dedicated tools described in detail by \citet{2015MNRAS.454..129V} and \citet{2016A&amp;A...587A..57Z}. We refer to these studies for a
complete description of the data reduction outlined below.

%_________________________________________________
\subsection{IFS pre-processing}
\label{sec:ifs_preprocessing}
The DRH was first used to create the basic files of the data reduction process: backgrounds, master flat-field, IFS spectra positions and
IFU flat-field following \citet{2015A&amp;A...576A.121M}. Four different lasers illuminate the IFS detector to allow for wavelength calibration
\citep[][]{2015A&amp;A...576A.121M, 2008SPIE.7019E..39P}. The dedicated pipeline presented by \citet{2015MNRAS.454..129V} was then
used to process the science frames. In a first stage, this pipeline performs the following steps: 1) determination of the time and parallactic
angle of each science frame, 2) background-subtraction, 3) normalization of the data by the value of the DIT, and  4) temporal binning
(blocks of two consecutive frames were averaged to reduce the total number of frames). 
These pre-processed binned frames were then corrected for bad pixels and cross-talk (spurious signal from adjacent lenslets on the IFS that can
contaminate the final images), before being fed into the DRH to interpolate the data spectrally and spatially. At this stage, we performed a second
round of bad-pixel cleaning, as the IFS detector is strongly affected by bad pixels and the amount of remaining bad pixels at the shorter wavelengths can
be significant. Bad pixels were identified with a sigma-clipping routine and then corrected with the IDL procedure {\sc maskinterp}\footnote{
https://physics.ucf.edu/~jh/ast/software.html}
which performs a bicubic pixel interpolation using neighbouring good pixels. These frames are then re-calibrated in wavelength
to  identify a reference channel for which the absolute wavelength is accurately known and to find the scaling factor between the spectral channels
\citep[see Sect. A2 in][]{2015MNRAS.454..129V}. Combining the uncertainties of these two separate steps yields an upper limit of 1.4 nm for the
wavelength calibration error.

%_________________________________________________
\subsection{IRDIS pre-processing}
\label{sec:ird_preprocessing}
Pre-processing of IRDIS data is much simpler than IFS data as no wavelength calibration is needed. We used our own IDL tools to
pre-process the data. The individual science frames were background subtracted and flat-field corrected. Bad pixels were identified
and corrected following the same method as with the IFS data. No temporal binning was applied to this dataset. 

%_________________________________________________
\subsection{Off-axis PSFs and centering frames}
\label{sec:off_axis_reference_psf}
The off-axis PSFs were acquired using the same instrumental setup as in the coronagraphic images. This dataset is calibrated and
pre-processed within the same procedure previously described. The centroid of each satellite spot in the centering frames was
determined by fitting a 2D Gaussian. The coordinates of the star over the detector were calculated as the center of these four
centrosymmetric spots. The accuracy of this procedure is estimated to be within a few tenths of a pixel \citep{2015A&amp;A...576A.121M}.
For the IFS, the star center is independently determined for each wavelength. The center of the PSF was very stable across the whole
observation, with a maximum displacement of $\sim0.2$ px in the X and Y directions on the detectors, and therefore no re-centering was applied.

%_________________________________________________
\subsection{Final image corrections}
\label{sec:final_corr}
After pre-processing, the images were corrected for anamorphism \citep[see][]{2016A&amp;A...587A..56M} using customized IDL
routines. We used the True North (TN), plate scales, and relative orientation between IRDIS and IFS derived by \citet{2016A&amp;A...587A..56M}:
plate scale of $7.46\pm 0.01$ mas px$^{-1}$ for the IFS, and $12.255 \pm0.012$ mas px$^{-1}$ for IRDIS, and a relative orientation
between IRDIS and IFS of $-100.46\degr \pm 0.13\degr$. Following these steps, the images were ready to be processed with dedicated
high-contrast algorithms described in the following section.

The sharp PSF provided by the adaptive optics (AO) system, combined with the effective central starlight suppression provided by the coronagraph,
made it possible to directly detect the light scattered by the disk at all wavelengths measured. As no comparison star was observed, and the disk is
nearly face-on \citep[$i\sim6\degr$,][]{2012ApJ...753...59M}, all the images were derotated to a common north and median combined without
applying any PSF-subtraction techniques. Higher signal to noise (S/N) was attained by combining all images in our dataset. The IFS channels were
combined to create broadband images at $Y$, $J$, and $H$ bands. Each pixel in the disk image was divided by the peak of the PSF previous exposure
time normalization to estimate the star-disk flux ratio.

%__________________________________________________________________________________________
\section{Speckle noise subtraction}
\label{sec:speckle_post}

Our large dataset is well suited to high-contrast algorithms devoted to detecting faint point sources using ADI \citep{2006ApJ...641..556M},
spectral differential imaging \citep[SDI,][]{1999PASP..111..587R} and/or a combination of both. We used principal component analysis
(PCA) following the Karhunen--Lo{\`e}ve Image Projection (KLIP) method \citep{2012ApJ...755L..28S, 2015ApJ...803...31P}, as well as
the matched locally optimized combination of images (MLOCI) algorithm \citep{2015A&amp;A...581A..24W}, to take advantage of the
large library of PSF images available. The IRDIS and IFS datasets were treated separately as they are naturally different: while the
IRDIS data contains narrow, dual-band images, each IFS datacube contains 39 images at different wavelengths. After initial inspection
by eye we removed several frames with very poor seeing and/or AO open loops (less than $5\%$ of the whole dataset). In the subsequent
analysis (described below), we also explored the effects of applying frame selection by disregarding the frames where the integrated flux
inside the AO corrected area remained below $3,5,7\times\sigma$ of the median. We find that the highest contrast is reached when using
all frames as input (i.e., not applying frame selection).

%____________________________________
\subsection{Dual-band imaging with IRDIS}
\label{sec:irdis_ps_img}
We separately analyze the $K1$ and $K2$ channels, as well as their difference (DBI). The data was processed
with three independent, different pipelines: using the PCA-based pipelines detailed in \citet{2014A&A...572A..85Z, 2016A&amp;A...587A..57Z}
and \citet{2016A&A...587A..55V}, and the MLOCI algorithm presented in \citet{2015A&amp;A...581A..24W}.
In the first approach, for each frame, a PCA library based on all the other frames from the sequence is constructed. This yields 140 PCA modes
that are calculated over the whole image from an inner radius of $0\farcs1$ up to $0\farcs9$. Different numbers of modes were subtracted from the images,
resulting in residuals quickly converging after using 50 modes. The MLOCI algorithm injects fake companions in different sectors over each frame
and then it constructs a reference image that maximizes the S/N of the recovered companions. This reference frame is created combining all the other
frames from the sequence with the LOCI algorithm \citep{2007ApJ...660..770L}. The $K2$ images were affected by strong background emission when
compared to the $K1$ images \citep[as noted by][]{2016A&amp;A...587A..56M}. 

%____________________________________
\subsection{Spectral imaging with IFS}
\label{sec:ifs_ps_img}
Here we followed the data reduction described by \citet{2015MNRAS.454..129V} where a thorough analysis of different data reduction 
approaches are used to maximize the contrast in SPHERE/IFS observations. These authors show that maximum contrast is reached
when combining SDI together with ADI exploiting the PCA. In this way the information contained in the spectral dimension is used to further
reduce the speckle-noise in the images. As with the IRDIS data, PCA is used to construct an optimum PSF from all the frames available
for each spectral channel. This process uses the spatial re-scaling to match the speckle pattern at all wavelengths, taking advantage of
the relatively large spectral coverage of the IFS. Hence, instrumental speckles can be subtracted resulting in a significant gain in absolute
contrast. In the case of this dataset, a total of 70 IFS data cubes were available, resulting in 2730 PCA modes (70 cubes each with 39 wavelengths).
For each IFS frame, we subtracted up to 50\% of the modes, in steps of five modes. All the data for a given number of subtracted modes were
spatially re-scaled back to their original size, de-rotated to a common north, and mean-combined to produce a final broad-band image. As with
the IRDIS dataset, the residuals converge after using one third of the PCA modes.

%__________________________________________________________________________________________
\section{Results}
\label{sec:results}

%____________________________________
\subsection{Detection limits and constraints on planetary mass}
\label{sec:det_limits}

\begin{figure}
  \centering \includegraphics[width=\columnwidth, trim = 0 40 50 60]{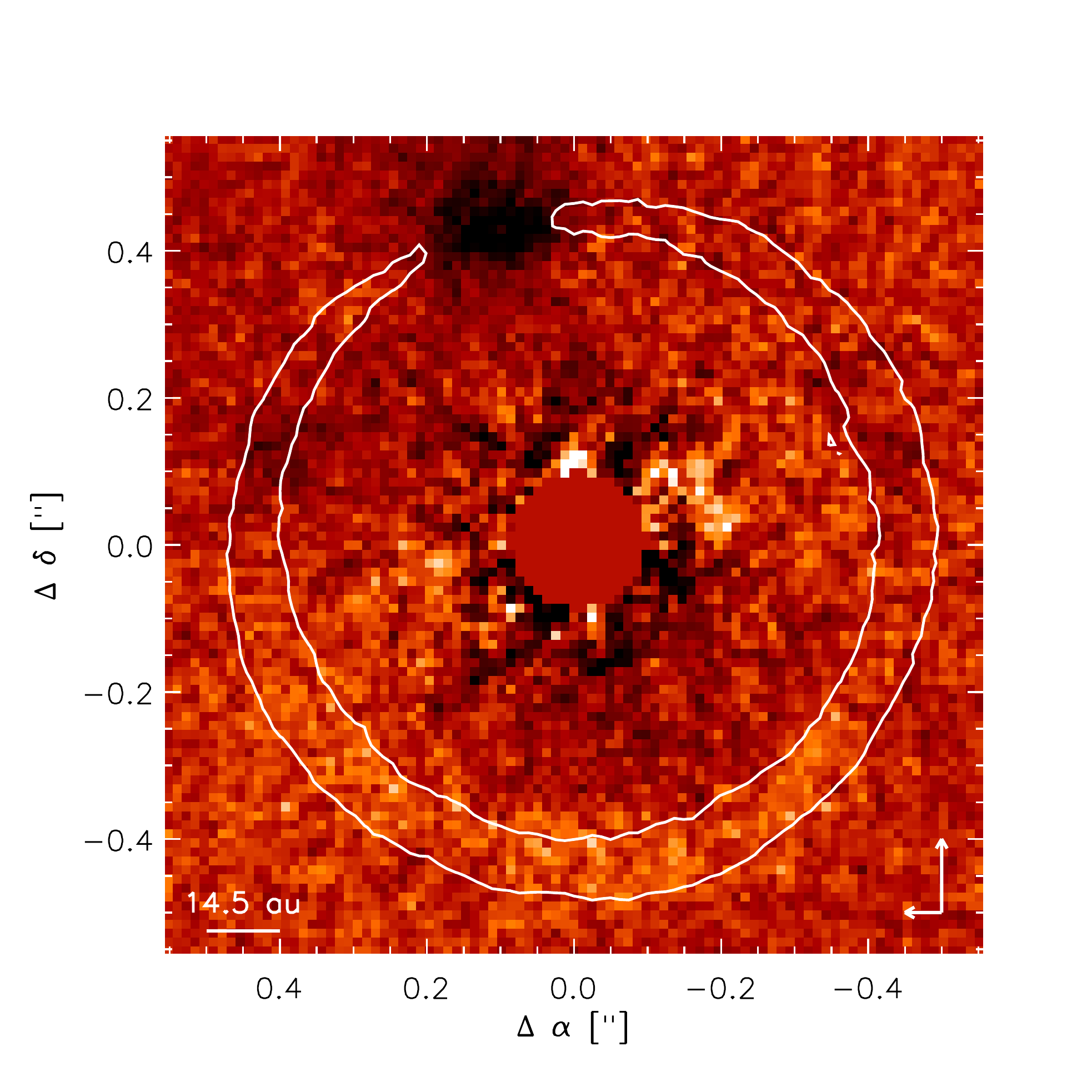}
  \caption{PCA reduced image of the IRDIS $K1$ filter. The image has a software mask of radii $0\farcs092$, equal to the IWA of our observations.
  The bright glow at the edge of the mask at PA $0\degr$ is an artefact. The dark region at $r \sim0\farcs4$ and PA $\sim18\degr$ corresponds to a
  dip in the disk bright inner rim. The white line contours the S/N = 30 region in the K1 image to highlight the disk morphology (see Sect.~\ref{sec:scattered_images}).}
   \label{fig:fig1}
\end{figure}

\begin{figure}
  \centering \includegraphics[width=\columnwidth, trim = 25 250 10 205]{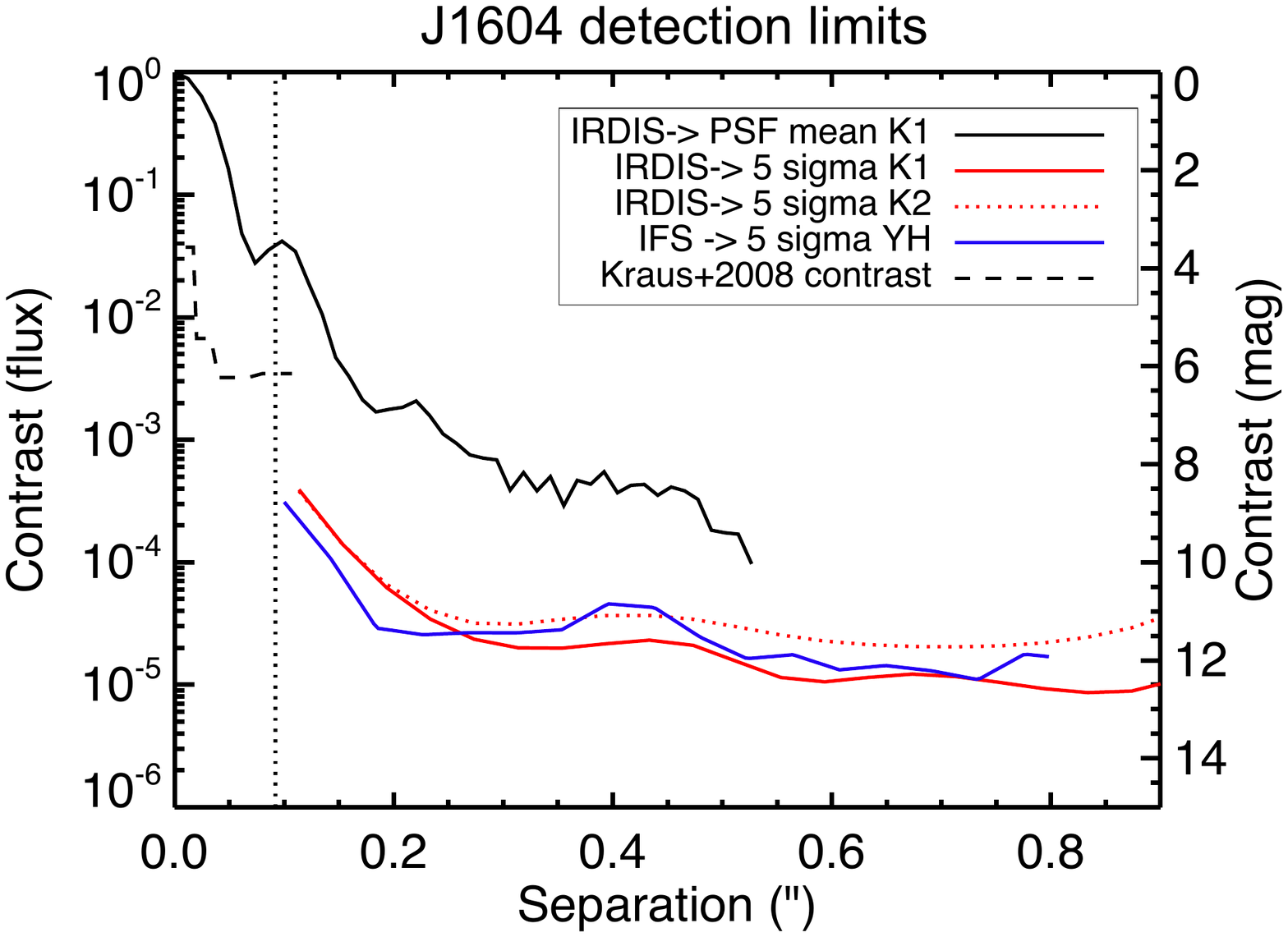}
  \caption{Azimuthally averaged contrast curves (5$\sigma$ detection limits) derived using the methods described in Sect.~\ref{sec:det_limits}.
  The black solid line traces the contrast of the averaged off-axis PSF at K1 band. The vertical dotted line indicates the IWA of our
  images. The dashed line indicates the detection limits derived by \citet{2008ApJ...679..762K}.}
   \label{fig:fig2}
\end{figure}

After careful inspection of the reduced images, no detection of a candidate planet was found in any of our datasets. For illustration we show
the PCA reduced image of the IRDIS $K1$ dataset in Fig.~\ref{fig:fig1}. The off-axis images were used to mimic faint companions to derive
detection limits from our observations. Twenty simulated planets were injected simultaneously in the pre-processed dataset at different radial
separations (starting at $0\farcs1$ and increasing in a $0\farcs04$ step) with respect to the star, at azimuthal angles spaced by $18\degr$.
This procedure was repeated 30 times rotating the position angles in steps of $12\degr$ each time to improve the statistical significance
of the results. The flux of the planets was scaled for seven different contrast levels ($[6, 2, 1]\times10^{-4}$, $[4,1]\times10^{-5}$, $[6,3]\times10^{-6}$)
with respect to the central star. The contrast was defined as the ratio between the integrated flux of the planet and that of the star. We assume
that the companions have the same spectra as J1604. As noted by \citet{2015MNRAS.454..129V}, this is not realistic, but has the advantage
of providing model-independent contrast limits that can be directly compared with other observations. We then applied the reduction steps
described in Sect.~\ref{sec:speckle_post}. A fake planet is considered as detected when its S/N is $\geq$5. The fluxes of the planets are
measured with aperture photometry using an aperture diameter of $0.8 \lambda /D$ (where $D$ is the telescope diameter). Self-subtraction
is estimated by comparing the planet flux before and after data processing. The measured flux is corrected by this effect and by the small
sample statistics at small radial separations \citep[$r<3\lambda/D$,][]{2014ApJ...792...97M}. Detection limits were calculated by measuring
the standard deviation of the final images in annuli of 1 $\lambda$/D width at increasing angular separation, normalized by the flux of the
average off-axis PSF at each band.
%=======================================================================
%=======================================================================
In IRDIS, the background emission in the $K2$ channel limits the detection threshold. By far the highest contrast was obtained when processing
the $K1$ channel alone. The detection limits from the IFS are similar to those from IRDIS in the $0\farcs5-0\farcs8$ range, and they are somewhat
lower in the inner regions, down to $0\farcs15$. The detection limits derived for IRDIS and for the IFS data are shown in Fig.~\ref{fig:fig2}.

%=======================================================================
%=======================================================================

\begin{figure}
  \centering \includegraphics[width=\columnwidth, trim = 0 30 0 0]{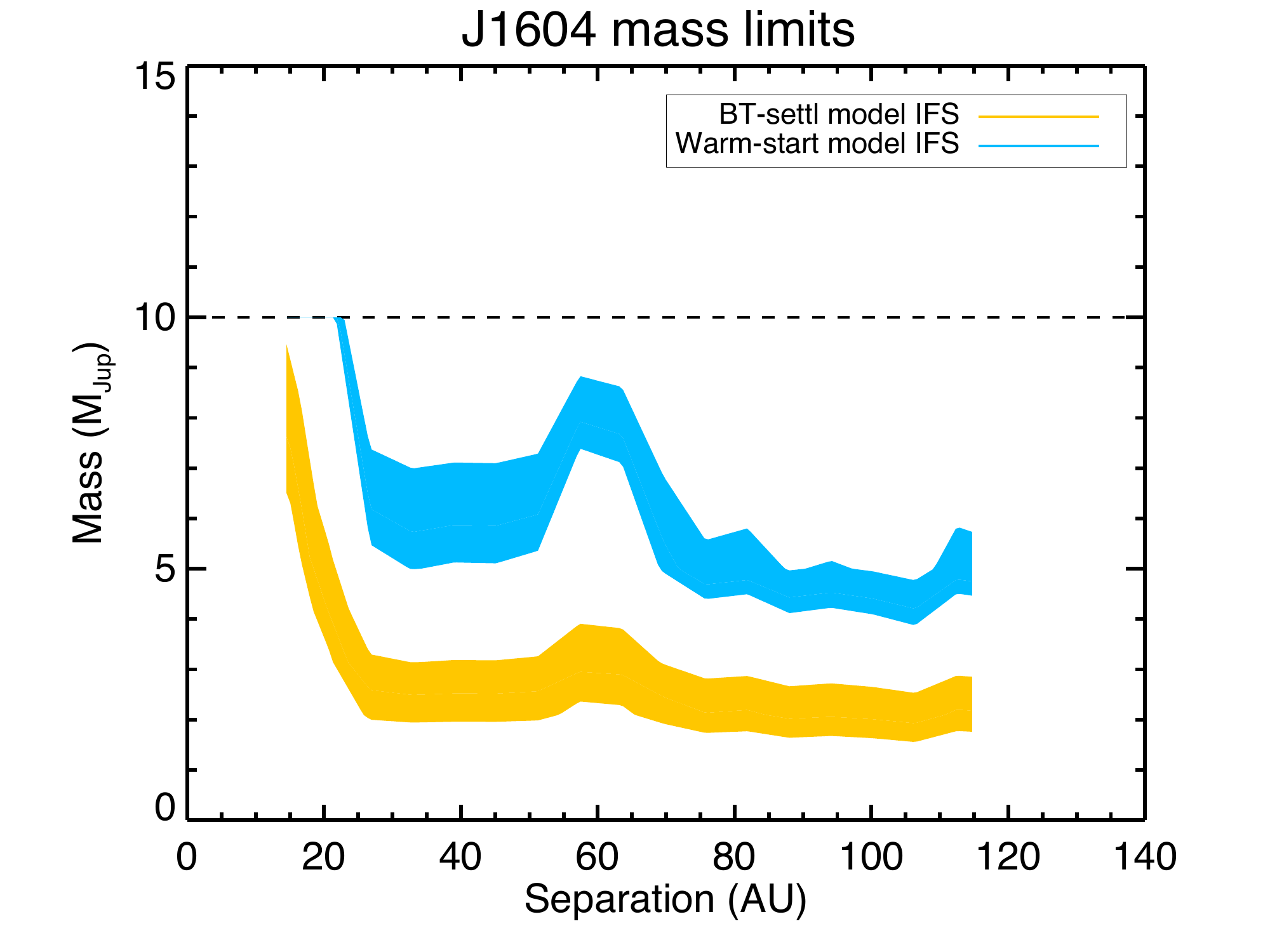}
  \caption{Physical limits on planetary mass companions derived as described in Sec.~\ref{sec:det_limits}. The IFS mass limits are slightly
  better than the IRDIS ones. The shadowed area reflects the large uncertainty on the age of the plausible companions, ranging from 5 to
  12 Myr. There are no warm start models for planets with masses above 10 $\MJup$. The `hump' in sensitivity at $\sim60$ au is due to
  disk emission. Mass limits from 115 to 130 au correspond to IRDIS K1 band observations using the same methodology as for the IFS.}
   \label{fig:fig3}
\end{figure}

Converting detection limits into planetary mass limits is not straightforward, as models studying the formation of young giant planets show
that the initial conditions of planet formation have a great impact on the planet luminosity \citep[e.g., ][]{2012ApJ...745..174S}. Here we followed
the method outlined by \citet{2016A&amp;A...587A..57Z} and \citet{2015MNRAS.454..129V}. We used two different evolutionary models
to estimate the flux of a young planet: the `hot start' model \citep[BT-Settl, ][]{2013MSAIS..24..128A, 2015A&amp;A...577A..42B} and the
`warm start' model with initial entropy of 9 $k_\mathrm{B} \, \mathrm{baryon^{-1}}$ and a cloudy atmosphere of one solar metallicity
\citep{2012ApJ...745..174S}. A major difference between these two models is the initial entropy of the planet embryo, which leads to
different planet fluxes. The models provide us with the absolute magnitude of the planet at the SPHERE bands\footnote{
The fluxes from \citet{2012ApJ...745..174S} were converted to the SPHERE photometric system using the filter transmission curves given
in https://www.eso.org/sci/facilities/paranal/instruments/sphere/inst/
filters.html}, 
which are then scaled to a distance
of 145 pc. Given the uncertainty surrounding the age of J1604, we derived the planet fluxes for a range of plausible ages from 5 to 12 Myr. These
values were compared to the J1604 flux at each band to estimate the model predicted contrast. The uncertainties on the distance and the
magnitude of the host star are negligible with respect to the other quantities on the calculation. Our mass detection limits were estimated by
comparing the predicted contrast with our detection limits. The results are shown in Fig.~\ref{fig:fig3}, where the shadowed areas illustrate
the spread in mass sensitivity at a given mass for the different ages, where the boundaries are 5 and 12 Myr. For simplicity we only show the
age spread for the mass limits derived from the IFS (which are lower than the IRDIS ones) up to $0\farcs8$. The IRDIS K1 mass limits are
shown in the region not reached by the IFS, from $0\farcs8$ to $0\farcs9$ (115 to 130 au). There are no warm start model fluxes for massive
planets ($>10\MJup$).     

%=======================================================================
%=======================================================================
%=======================================================================
\begin{figure*}[!t]
  \centering \includegraphics[width=\textwidth, trim = 0 40 0 30]{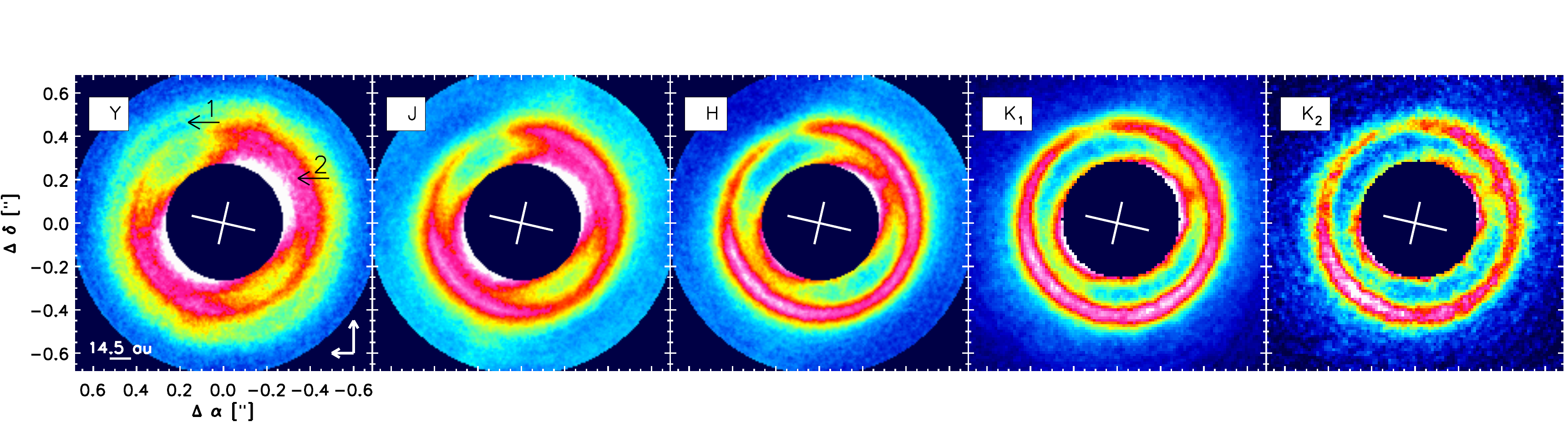}
  \caption{ Scattered light images of the disk (Sect.~\ref{sec:final_corr}) at different bands after applying an unsharp filter as described in
  Sect.~\ref{sec:scattered_images}. A software mask of radius $r = 0\farcs27$ is used to remove the spurious signal that dominates the central
  regions of the images. The dynamical range of each image has been arbitrarily scaled to highlight the disk features, and the color scale is linear.
  The cross at the center indicates the major/minor axis of the disk, at PA of $76.8\degr$ \citep{2014ApJ...791...42Z}. Arrow `1' points to the faint,
  ring-like halo artefact observed at $r\sim0\farcs5$ at $J$ band, while arrow `2' indicates the elliptical artefact that especially affects the IFS
  images, with increasing strength towards shorter wavelengths. In all images, a dip/decrement in brightness at $\sim20\degr$ (east form north)
  is evident in the bright inner rim of the disk, although it is partially masked by artefact `2' (see text). This dip was previously noticed by
  \citet{2015A&amp;A...584L...4P} at $R'$ band.}
   \label{fig:fig4}
\end{figure*}
%=======================================================================
%____________________________________
\subsection{Scattered light images of the disk}
\label{sec:scattered_images}

The disk images are affected by two major artefacts. First, we find a faint, ring-like halo of remnant speckles. This annular structure increases
linearly in radius with wavelengths starting at $r\sim0\farcs5$ at $J$ band. Second, we also find a large-scale, elliptical artefact with its major
axis aligned at roughly $140\degr$ east of north and centered behind the coronagraph. This is most likely created by diffraction at the edges
of the mask and non-perfect AO correction (probably due to high-altitude winds) during the observations. A similar artefact also appears
in the SPHERE dataset presented by \citet{2016A&amp;A...587A..56M} (see their figures 1 and 2), albeit with much lower amplitude. At shorter
wavelengths, this artefact dominates the disk emission, and our best images (higher S/N and less affected by artefacts) correspond to the $H$
and $K1$ datasets. Applying frame selection does not significantly reduce the contamination. In an attempt to mitigate this effect and highlight
morphological structures in the disk we applied an unsharp mask with a kernel of $2\times$FWHM at each band to the images. The unsharpened
images are shown in Fig.~\ref{fig:fig4}. A dip or decrement in scattered light at $\sim20\degr$ (east from north) is detected in each band and
especially at $K1$. 

We created a toy model of the disk using the radiative transfer code MCFOST \citep{2006A&amp;A...459..797P, 2009A&A...498..967P}
aiming to estimate the effect of the elliptical artefact over the disk signal. The dust population is made by porous grains composed of
$80\%$ astronomical silicates \citep{1984ApJ...285...89D} and $20\%$  carbonaceous amorphous particles \citep{2003ARA&amp;A..41..241D}
following a standard power-law size distribution d$n(a)\propto a^{-3.5}$d$a$ with grain sizes ranging from $a_{\mathrm{min, max}} = 0.05\mu$m to 1mm. 
The full scattering matrix of the dust population is computed with Mie theory \citep{mie_1908} using the distribution of hollow
spheres (DHS) formalism outlined by \citep{2005A&A...432..909M}, with maximum volume fraction $f_\mathrm{max}$ = 0.8.
The surface density distribution $\Sigma(r)$ is described by a standard power-law with a tapered-edge profile
% EQ 1================
\begin{equation}
\Sigma (r) = \Sigma_{\rm{C}} \, r^{-\gamma} \, \mathrm{exp} \left [ - \left (\frac{r}{R_{\rm{C}}} \right) ^{2-\gamma} \right ],
\label{eq:eq1}
\end{equation}
% ====================
where $r$ is the radial distance from the star, $R_\mathrm{C}$ is the characteristic radius $R_\mathrm{C} = 100$ au, $\Sigma_{\rm{C}}$ is the surface density at $R_\mathrm{C}$,
and $\gamma = 1$. We do not consider the innermost (observationally poorly unconstrained) material here. The inner $r = 61$ au region is
totally depleted of dust, and the sharp outer edge of this cavity is smoothed with a Gaussian taper of 2 au FWHM. The vertical structure of
the disk is assumed to follow a Gaussian density profile, and the scale height at each radius is then defined as $H(r) = H_{100} (r / 100\,\mathrm{au})^{\psi}$,
where $\psi$ defines the flaring angle of the disk and $H_{100}$ is the characteristic scale height at 100 au. We use the stellar parameters,
distance, disk position angle and inclination given in Sect.~\ref{sec:intro}. Scattered light images of the disk at the bands of our observations
are created with a ray tracing algorithm. These images are convolved with the corresponding off-axis real observations. The signal from the disk
in these non-coronagraphic PSFs has a contrast of $\sim10^{-4}$ with respect to the star peak and therefore its contribution is negligible.
Gaussian noise is finally added to these images. The synthetic image at $H$ band is shown in the left panel of Fig.~\ref{fig:fig5}. In our model, the near side
corresponds to the south-east side of the disk. The scattered light along the projected major axis is symmetric, whereas along the minor axis
the emission is asymmetric with the nearest side $\sim30\%$ brighter than the farther one.

We then assume that the large scale artefact could be described by a broad elliptical gaussian with its major axis aligned at $140\degr$
east of north and a major/minor axis ratio of $\sim1.2$. This synthetic artefact is then added to the MCFOST images, and it is arbitrarily
scaled until it roughly matches our observations (see Fig.~\ref{fig:fig5}). 
With the exception of the dip located at $\sim20\degr$, the synthetic images reproduce the observed overall azimuthal brightness modulation:
a decrement in the $[-20\degr, 60\degr]$ and $[170\degr,250\degr]$ ranges, and maxima at $\sim130\degr$ and  $\sim290\degr$, (east of north).
Therefore we consider it very likely that the azimuthal brightness modulation is a consequence of the large scale artifact, which masks
real disk features especially at $H$ band and shorter wavelengths. We emphasize that the dip is a real astrophysical feature that has been
previously reported using different observations \citep{2015A&amp;A...584L...4P}. As a separate test, we repeat the previous exercise but
with the near side of the disk on the north-west. In this case the final images do not reproduce the overall observed morphology.

%____________________________________
\subsubsection{Brightness radial profile}

Having qualitatively identified the major effects of the artefacts contaminating our images, we proceed to estimate the radial brightness profile
of the disk for the $H$ and $K1$ band images. To that end we computed the median and standard deviation at each position in a 3-pixel width
slit along the major and minor axis of the disk. At $K1,$ the disk is detected up to $\sim0\farcs72$ or $\sim104$ au. The results for the major
axis are shown in Fig.~\ref{fig:fig6}. In both bands the western side is brighter than the eastern one, and this difference is smaller for the $K1$
images (as it is less contaminated by the artefacts). We consider it most likely that this difference is due to contamination by the elliptical artefact
described above, as the brightness profile is expected to be centrosymmetric along the major axis of an inclined disk. In all cases, the profiles
peak at $0\farcs43$ ($\sim63$ au), in agreement with previous findings \citep{2012ApJ...760L..26M, 2015A&amp;A...584L...4P}.

The radial brightness profile of a protoplanetary disk is expected to decrease with radius following a power law \citep[$\propto r^{\,\alpha}$,][]
{1992ApJ...395..529W}. We use a Levenberg-Marquardt algorithm giving weights to each point according to their standard deviations to fit
power-laws to the measured profiles. This way we find that a better fit (lower reduced $\chi^{2}$) is attained when splitting each fit into two
separate regions. We define the radius at which the power-law changes as the `breaking point' ($r_{bp}$). The different fits for each band
are listed in Table~\ref{tab:tab1} and given in the legend of Fig.~\ref{fig:fig6}, following the same color code as the plots in the figure. Overall
we find that the profiles on the eastern side are slightly more shallow than their counterparts on the western side, albeit this difference being within
$2\sigma$. The breaking point $r_{bp}$ remains almost constant for the two bands and the two sides (see Table~\ref{tab:tab1}), which suggest
that this is a real feature and not a consequence of artefact contamination. Averaging the absolute values of $r_{bp}$ we obtain
$r_{bp} = 0\farcs55\pm0\farcs03$ ($79.75\pm4.35$ au at 145 pc). Interestingly, this value matches the outer edge of the gap of the mm-sized
dust distribution derived from ALMA observations \citep[$\sim79$ au, ][]{2014ApJ...791...42Z}.

%=======================================================================
\begin{figure}[!t]
  \centering \includegraphics[width=\columnwidth, trim = 0 20 0 0]{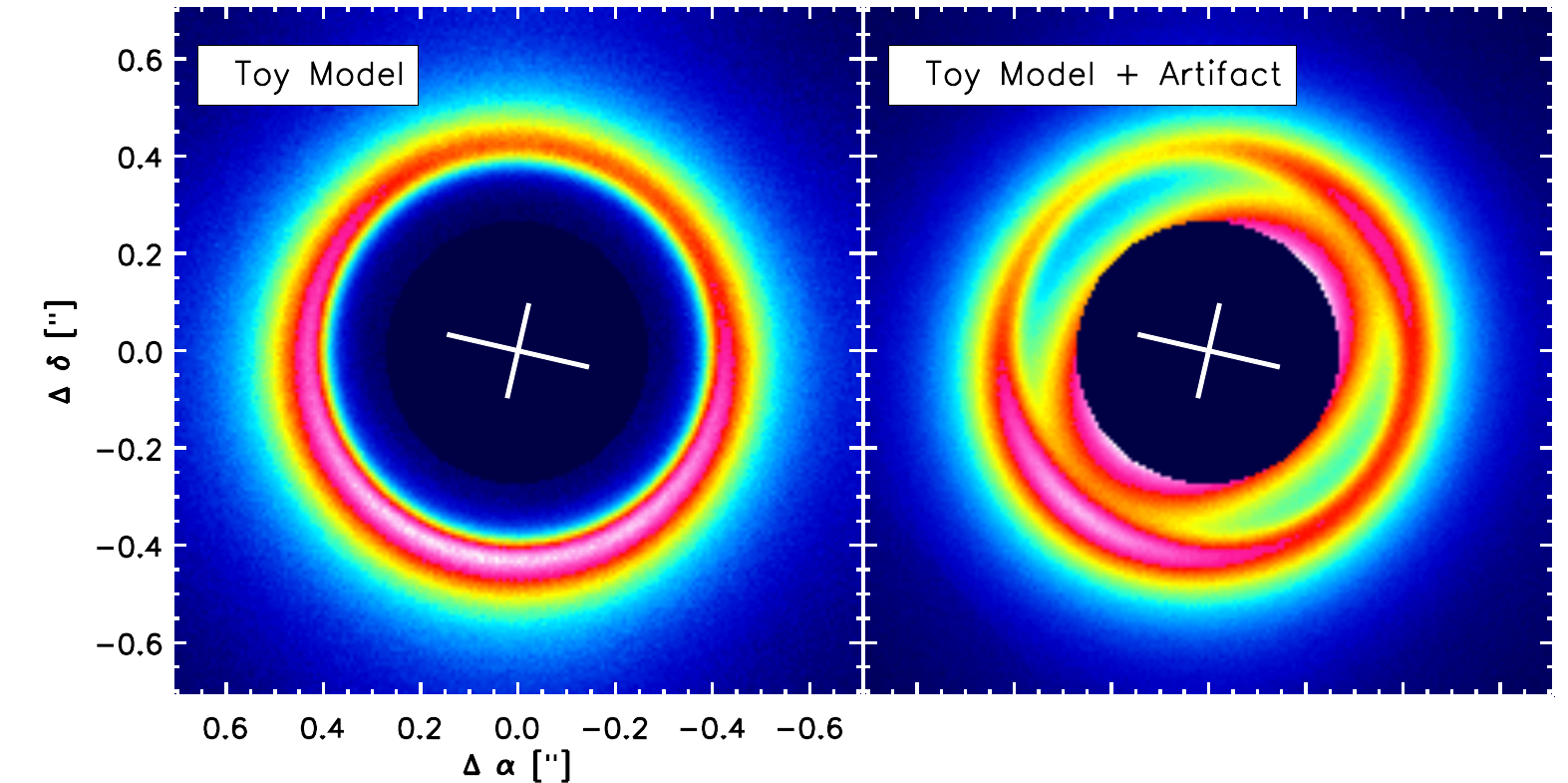}
  \caption{\textbf{Left:} Synthetic $H$ band image of the toy model presented in Sect.~\ref{sec:scattered_images}. \textbf{Right:} Same
  model presented in the left panel, but adding an elliptical gaussian artefact similar to that affecting our observations (see text for details).}
  \label{fig:fig5}
\end{figure}
%=======================================================================

As the disk is nearly face-on, we repeated the previous exercise but this time computed the profiles along the minor axis (due to projection effects,
the difference between the minor and major axis remains below $0.5\%$). The southern side is brighter than the northern one (Fig.~\ref{fig:fig6}),
as expected for an inclined disk with its nearest side on the southern direction (Fig.~\ref{fig:fig5}, left panel). This difference is higher (up to $\sim15\%$)
at $K1$ band. As it happens, with the major axis, we find that using a broken power-law produces a much more robust fit. The results from this fit
are given in Table~\ref{tab:tab2}. We find again that the absolute value of the breaking point remains roughly constant for all the fits, and in excellent
agreement with its counterpart along the major axis, with an average value of $r_{bp} = 0\farcs54\pm0\farcs01$ ($78.3\pm1.5$ au).

%=======================================================================
\begin{figure*}[!t]
  \centering \includegraphics[width=\textwidth, trim = 60 20 90 0]{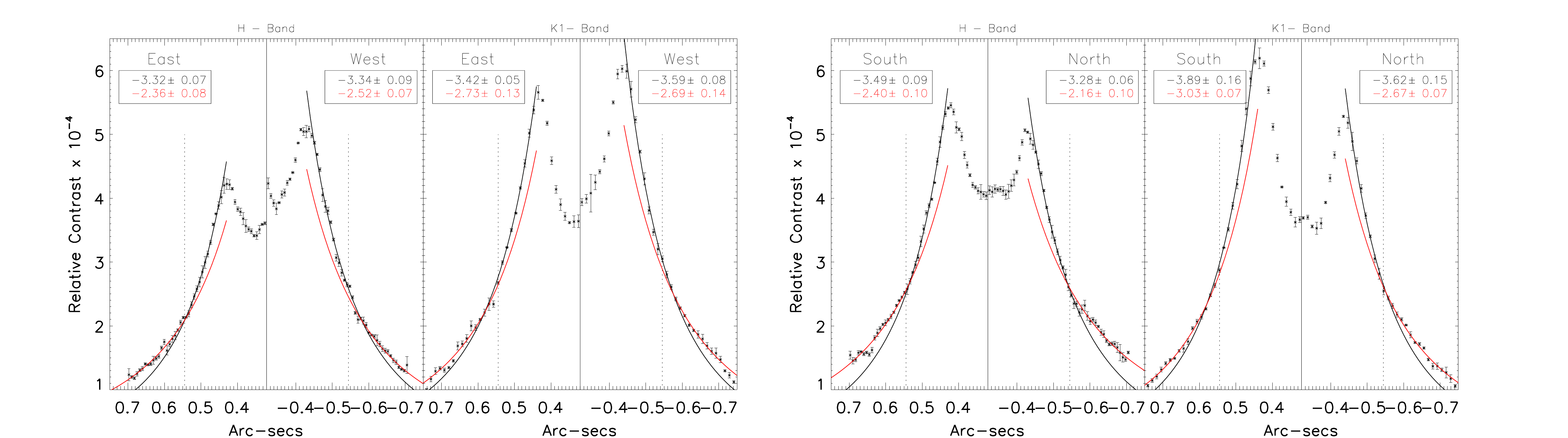}
  \caption{Cuts along the disk major (left panel) and minor (right panel) axis. Each panel shows the cuts at $H$ and $K1$ bands. Black asterisks and error bars
  represent the median and standard deviation at each position over a 3-px width slit along the major axis. A broken power-law profile is fitted at each side of
  the profile. The power-law indices are given in the legends of each panel, with the same color code as the fits plotted as solid curves, and correspond to the
  values listed in Table~\ref{tab:tab1} (major axis) and Table~\ref{tab:tab2} (minor axis). The vertical dotted line indicates the $\sim79$ au cavity in the mm-sized
  grains \citep{2014ApJ...791...42Z}. The cuts are given in contrast (units of $10^{-4}$) with respect to the star at each band.}
   \label{fig:fig6}
\end{figure*}
%=======================================================================

%_______________________________________________
%__ Table 1 ______________________________________
\begin{table}[t]
\center
\caption{Broken power-law ($\propto r^{\,\alpha}$) fits to the radial brightness profile along the major axis shown in Fig.~\ref{fig:fig6}. The errors
represent the $1\sigma$ uncertainty from our fit.}
\begin{tabular}{ c  c  c c}  
\hline\hline
Side      & Band    &   Range                                &  Power-law ($\alpha$) \\
\hline
East      &  $H$     &  $-0\farcs44 : -0\farcs53$    &  $-3.32\pm0.07$   \\
East      &  $H$     &  $-0\farcs53 : -0\farcs70$    &  $-2.36\pm0.08$   \\
West     &  $H$     &  $ 0\farcs44 :  0\farcs56$    &  $-3.34\pm0.09$   \\
West     &  $H$     &  $ 0\farcs56 :  0\farcs70$    &  $-2.52\pm0.07$   \\
\hline
East      &  $K1$   &  $-0\farcs44 : -0\farcs54$    &  $-3.42\pm0.05$   \\
East      &  $K1$   &  $-0\farcs54 : -0\farcs72$    &  $-2.73\pm0.13$   \\
West     &  $K1$   &  $ 0\farcs44 :  0\farcs59$    &  $-3.59\pm0.08$   \\
West     &  $K1$   &  $ 0\farcs59 :  0\farcs72$    &  $-2.69\pm0.14$   \\
\hline
\end{tabular}
%               \tablefoot{}
\label{tab:tab1}
       %\tablefoottext{}{} 
\end{table}
%_______________________________________________

%_______________________________________________
%__ Table 2 ______________________________________
\begin{table}[t]
\center
\caption{Same as Table~\ref{tab:tab1}, but along the minor axis.}
\begin{tabular}{ c  c  c c}  
\hline\hline
Side      & Band    &   Range                                &  Power-law ($\alpha$) \\
\hline
South    &  $H$     &  $-0\farcs44 : -0\farcs53$    &  $-3.49\pm0.09$   \\
South    &  $H$     &  $-0\farcs53 : -0\farcs70$    &  $-2.40\pm0.10$   \\
North     &  $H$     &  $ 0\farcs44 :  0\farcs56$    &  $-3.28\pm0.06$   \\
North     &  $H$     &  $ 0\farcs56 :  0\farcs70$    &  $-2.16\pm0.10$   \\
\hline
South    &  $K1$   &  $-0\farcs44 : -0\farcs54$    &  $-3.89\pm0.16$   \\
South    &  $K1$   &  $-0\farcs54 : -0\farcs72$    &  $-3.03\pm0.07$   \\
North     &  $K1$   &  $ 0\farcs44 :  0\farcs54$    &  $-3.62\pm0.15$   \\
North     &  $K1$   &  $ 0\farcs54 :  0\farcs72$    &  $-2.67\pm0.07$  \\
\hline
\end{tabular}
\label{tab:tab2}
\end{table}
%_______________________________________________

%____________________________________
\subsubsection{Azimuthal brightness profile}
\label{sec:azim:pro}

\begin{figure}[!t]
  \centering \includegraphics[width=\columnwidth, trim = 30 30 10 0]{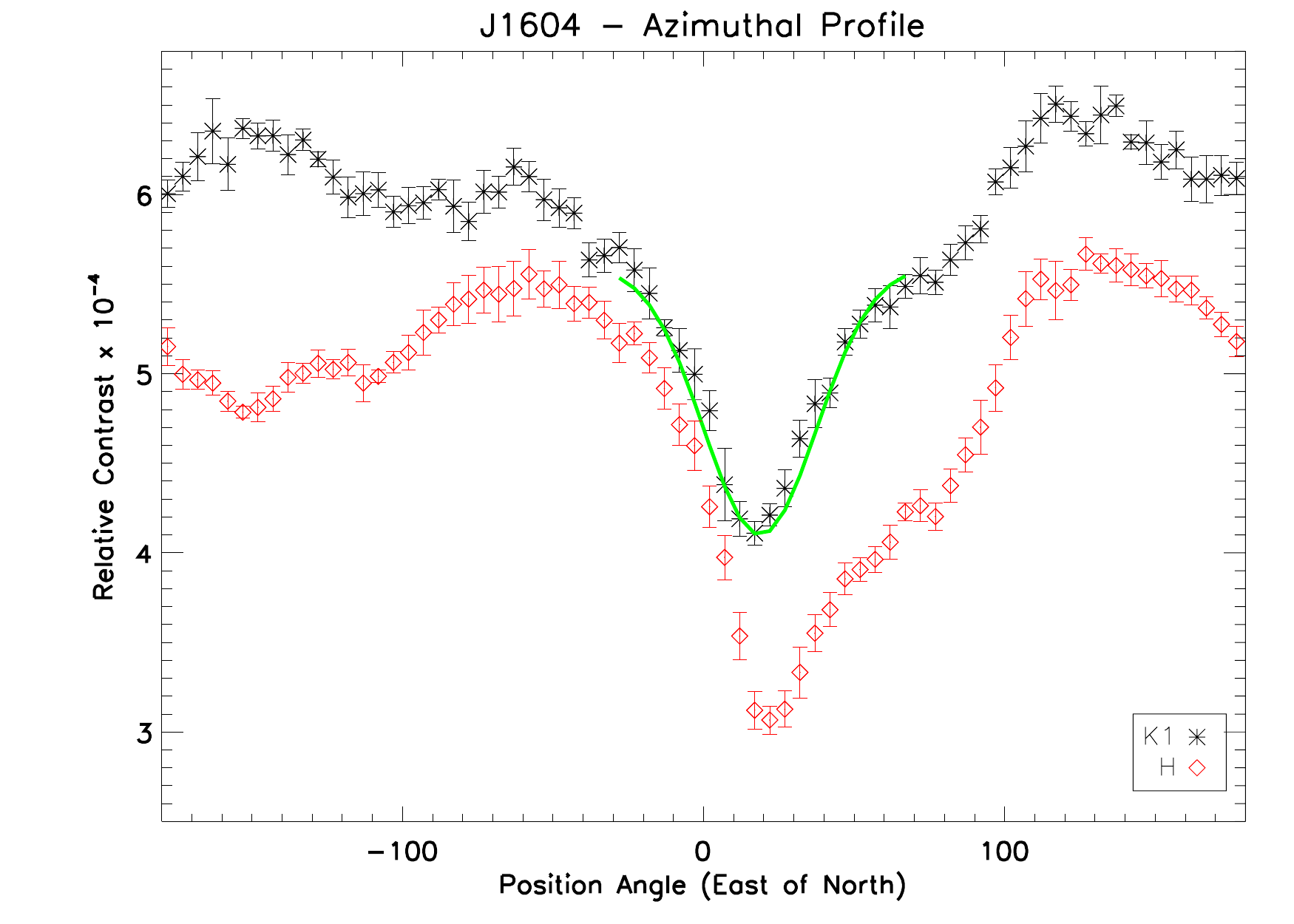}
  \caption{Azimuthal profile at $H$ and $K1$ bands, computed in a 2-px-width ring tracing the maximum of the ring at $\sim0\farcs43$ ($\sim62$ au at 145 pc).
  Each point is the median and its corresponding $1\sigma$ in a $5\degr$ step. The green line shows a Gaussian fit to the $K1$ profile (see text).}
   \label{fig:fig7}
\end{figure}

The azimuthal brightness profile was computed along the bright rim of the disk at $H$ and $K1$ bands. The median and standard deviation
were computed along the azimuthal direction using a 2-px width wedge with $5\degr$ opening angle (Fig.~\ref{fig:fig7}). The apparent local
maxima around $\sim130\degr$ and $\sim-50\degr$ are aligned with the major axis of the elliptical artefact, thus they are probably a consequence
of this artefact contamination. Excluding the artefact-dominated emission, the decrement 
in the dip with respect to the median of the $-180\degr:-100\degr$ and $100\degr:180\degr$ regions is $\delta_{K1} \sim 0.7$ and $\delta_{H} \sim 0.6$.
The region between $-60\degr$ and $100\degr$ (i.e., the northern side) shows lower brightness, as expected for an inclined disk with its nearest side on the south. 
We performed a non-linear, least squares Gaussian fit to the azimuthal profiles in the $-30\degr$ to $  70\degr$ range. The $K1$ fit produces a
reduced Chi-squared of  $\tilde{\chi}^2 = 0.9$ with the center of the Gaussian located at $18.8\degr \pm 0.7\degr$. The $H$ band profile fit does
not converge, most likely due to its higher contamination by artifacts.

%=======================================================================
%=======================================================================

%__________________________________________________________________________________________
\section{Discussion}
\label{sec:discussion}

As explained in the introduction, the radial structure of J1604 (both in dust and gas) matches several predictions by models of planet-disc interaction.
Additionally, the accretion rate of J1604 is remarkably low
\citep[$\dot{M}_{acc} <10^{-11} \MSun$ yr$^{-1}$; EW (H$_{\alpha}) = 5.3\AA$,][]{2012ApJ...753...59M, 2016A&amp;A...592A.126V} compared
to the typical values found for most pre-transitional disks \citep[median $\dot{M}_{acc} = 10^{-8.25} \MSun $yr$^{-1}$,][]
{2013ApJ...769..149K}. This is surprising considering the substantial amount of gas still present on the outer disk \citep[$M_{gas} = 0.02\,\MSun$][]
{2015A&amp;A...579A.106V} despite the relative old age of the system (5-12 Myr).

The complex radial structure of the disk, and in particular the hot dust inside of the gas cavity, disagrees with photoevaporation or
grain growth as potential mechanisms carving the disk \citep[as already noted by][]{2012ApJ...745...23M, 2014ApJ...791...42Z, 2015A&amp;A...584L...4P}.
Dead-zones \citep[poorly ionized disk regions, see e.g.,][]{1996ApJ...457..355G, 2006A&A...446L..13V, 2012MNRAS.419.1701R}
can create ring-like accumulations of large grains as a consequence of the change in viscosity at the outer edge of the dead-zone.
However this mechanism does not create deep cavities in the gaseous distribution, and recent models predict a second ring composed
by large grains also in the inner edge of the dead-zone \citep[][]{2016A&amp;A...590A..17R}. Combining our results with the
aperture-masking results by \citet{2008ApJ...679..762K}, stellar-mass companions are ruled out down to 2 au. In the most favorable
case of an equal mass companion following an elliptical orbit with semi-major axis of 2 au, tidal truncation could at most open a $\sim5$
au (in radius) cavity \citep{1994ApJ...421..651A} in the circumbinary disk. Therefore binarity cannot explain the radial structure of J1604.
To summarise, the structure of J1604 most likely reflects interactions between the disk and sub-stellar companion(s). 

According to the BT-Settl models (hot start scenario), our observations are sensitive to planets with masses exceeding $\gtrsim2.5\MJup$
across the disk cavity, except in the 14.5-22 au range where the sensitivity drops to $\gtrsim 9-4$ $\MJup$ (Fig.~\ref{fig:fig3}). The warm start
scenario is less stringent, and predicts that planets below $\sim6\MJup$ will escape detection. Some studies associate the hot- and warm-
start scenarios to planet formation via gravitational instability and core accretion, respectively. However, \citet{2013A&amp;A...558A.113M}
show that care must be taken with this classification as planets formed via core accretion may have similar properties to those formed via
a hot start scenario.

%____________________________________
\subsection{Multiple Jovian planets?}
\label{sec:discussion_1}
Planets and disk properties can be related by the widths and depths of planetary gaps produced by numerical simulations.
The size of the cavity carved by a planet depends on the planet orbit ($r_p$) and the Hill radius $r_{H} = r_p (q/3)^{1/3}$,
where $q$ is the planet/star mass ratio. \citet{2016MNRAS.459L..85D} show that large cavities carved by planets in the
mm- and $\mu$m -sized grain populations most likely indicate a turbulent mixing (directly related to the disk viscosity) of
$\alpha_{turb}\sim10^{-3}$. In what follows, we adopt this value for comparison with model predictions. Following \citet{2012A&amp;A...545A..81P},
one massive ($\ge 5\MJup$) giant planet will open a gap of $\sim5 r_H$ in the gas and $\sim10 r_H$ in the dust,
that is, the dust cavity can be up to twice the size of the gas cavity. \citet{2015ApJ...809...93D} analyzes the effect of multiple
planets, finding that multiple low-mass planets (e.g., $3\times0.2\MJup$) cannot open a large dust gap/cavity as they will just clear
up narrow gaps. Regarding the depth of the cavity, planets reduce the amount of gas inside of the cavities, with more massive
planets creating more gas-depleted cavities. For instance, one $9\MJup$ planet should deplete the gas surface
density by a factor $\sim10^{-3}$, which is roughly two orders of magnitude higher than what is observed for J1604
\citep[$\delta_\mathrm{co}\sim 10^{-5}$,][]{2015A&amp;A...579A.106V}. On the other hand, \citet{2015ApJ...802...42D} show
that, in general, gas cavities created by multiple planets are shallower (i.e., less gas depleted) than those created by
one single planet. Finally,  planets can also reduce the accretion rates onto their host star \citep[e.g.,][]{2007MNRAS.375..500A}, and
\citet{2011ApJ...729...47Z} show that for a standard disk accreting at $\dot{M}_{acc} \sim 10^{-8} \MSun $yr$^{-1}$, multiple
planets are needed to reduce the accretion rate below $\dot{M}_{acc} = 10^{-9} \MSun $yr$^{-1}$ \citep[see also][]{2014prpl.conf..497E}.

Combining all the previous predictions, one configuration in agreement with our observations and current theories of planet-disk interaction
could be the following: 
a $\sim15\MJup$ brown dwarf orbiting at $\sim15$ au (or more massive at closer orbits) would remain undetected, while it could account for the $\sim30$ au,
$\delta_\mathrm{co}\sim10^{-5}$ gas depleted cavity \citep{2012A&amp;A...545A..81P, 2014ApJ...782...88F}.
One or more Jovian planets at larger orbits could then explain the very low accretion rate and larger dust cavities, while creating
very narrow and shallow gaps in the gas that would remain undetected with the sensitivity and spatial resolution of current ALMA
observations. Of course, the scenario we just described is only one possibility. In general, and despite our deep observations, 
this problem is degenerated, which prevents us from deriving further constraints on the orbits and masses of this hypothetical
multi-planetary system.

%____________________________________
\subsection{Disk surface}
\label{sec:discussion_2}

The radial brightness profiles along the major and minor axes (Fig.~\ref{fig:fig6} and \ref{fig:fig7}) show that the dust in the disk surface scatters
slightly more light towards longer wavelengths. We note that these profiles are given in contrast units for each band, such that the variation of stellar flux
with wavelength is automatically considered. Taking into account that the $H$ band image is more contaminated by the elliptical artefact (i.e., receives
more flux from it) than the $K1$ image, then the brightness profiles indicate that the disk has a {red} color. Scattering in the Rayleigh regime
($2\pi a < \lambda$) produces very blue colours (as the scattering efficiency in this regime scales with $\lambda^{-4}$). Therefore, our images indicate
that the population of grains on the disk surface is dominated by grains with an average size $\gtrsim0.3\,\mu m$ \citep[see e.g., ][for a discussion
on dust properties derived from scattered light images at NIR]{2013A&amp;A...549A.112M}. The artefact contamination makes it difficult to obtain a
more detailed description of the dust properties. Imaging polarimetry observations of other planet-forming candidate disks also reveal broken-power
law brightness profiles \citep[HD\,169142, HD\,135344B,][]{2013ApJ...766L...2Q, 2013A&amp;A...560A.105G} and red disk colours
\citep[HD\,100546, HD\,142527,][]{2013A&amp;A...549A.112M,2013A&amp;A...556A.123C,2014ApJ...781...87A}.

There are different estimates for the power-law index of the radial profile in the literature, all derived from polarized intensity observations.
\citet{2015A&amp;A...584L...4P} computes the {azimuthally} averaged radial profile at $R$ band, obtaining a power-law index
of $\alpha = -2.92\pm0.03$. \citet{2012ApJ...760L..26M} follow a different approach and fit a power-law along the major axis using a
$30\degr$ width slit centered at $80\degr$ east of north, obtaining $\alpha = -4.7\pm0.1$ and $\alpha = -4.0\pm0.2$ for the  the eastern
and western sides, respectively. While these values are different from those presented here, we note that they were computed over a much
larger azimuthal range than our values listed in Table~\ref{tab:tab1}, which were computed using a narrow 3-px width slit centered on the
most recent value derived for the position angle ($PA \sim 77\degr$). The brightness profiles of our images become more steep
\citep[with power-law index $>-3$, see][]{1992ApJ...395..529W} at $\sim79$ au, which coincides with the cavity size for the large grains
derived by \citet{2014ApJ...791...42Z}. Inwards of that radius, our profiles become more flat (power-law index $<-3$). This suggests
that the surface of the disk, where the small grains are located, is sensitive to the change in the radial distribution of the large grains,
which are mostly concentrated around the disk's cold midplane. 

The observations presented by \citet{2015A&amp;A...584L...4P} were recorded one night later than the
data set we present here. They fit the dip with a Gaussian profile and derive a position angle of $46.2\degr \pm 5.4\degr$ east of north, and a depletion factor 
$\delta_\mathrm{dip}\sim0.72$. The relatively large difference between their derived position angle and ours ($\sim18.8\degr\pm0.6\degr$
at $K1$) is probably best explained by the much lower S/N of their images. Following \citet{2015A&amp;A...584L...4P}, assuming
that the dip observed by \citet{2012ApJ...760L..26M} is the same feature observed here, then it rotates at $\sim22\degr$ yr$^{-1}$
(e.g., period of $\sim16.4$ yr, roughly half the period previously derived). The dip could be a shadow casted
by an inner body/dust structure, located at a Keplerian orbit with
a $\sim6.3$ au radius. We note that a tilted inner disk should produce either two sharp shadows or an azimuthally large, smooth, single shadow
\citep{2015ApJ...798L..44M, 2016arXiv160300481S}. On the other hand, a point source, or a very flat structure, cannot cast a shadow
on the disk surface, as this shadow would be cast around the disk midplane only. We hypothesize that a low-mass companion surrounded
by a disk orbiting at $\sim6$ au, with a disk diameter of $\sim3$ au and scale height $\sim0.8$ au could cast a shadow over the disk surface.
This would create a dip in brightness with similar morphology to what is observed in J1604.

%__________________________________________________________________________________________
\section{Summary}
\label{sec:conclussions}
We present deep, high-contrast SPHERE/VLT observations of the pre-transitional disk J1604. This disk shows several
features that suggest planet-disk interactions, such as different cavity sizes in its gas and dust distributions. Despite
reaching a contrast of $\sim \Delta 12$ mag at $YJH$ and $K$ bands from $0\farcs15 - 0\farcs8$, we do not detect
any point source. Translating our limits to mass sensitivity, and assuming a distance of 145 pc and an age ranging from
5 to 12 Myr, we are sensitive to planets with masses $\gtrsim 2.5 \MJup$ if formed via a hot start scenario at $22$ to $115$ au
from the star. Alternatively, we are sensitive to $\gtrsim 6 \MJup$ in the same distance range if the planets formed via a
warm start scenario. Combining our non-detection with the observed cavity sizes and accretion rates of J1604, and comparing
these observational constraints with the predictions of models describing planet-disk interactions, we conclude that J1604 is
most likely a multi-planetary system. 

We find that the decrement in the disk bright inner rim is located at $\sim18.8\degr$ at $K1$ band. This dip could be produced
by a large structure in Keplerian motion around the central star. A consistent speculative scenario would thus be a brown dwarf
\citep[up to $20\MJup$ given the detection limits from ][]{2008ApJ...679..762K} surrounded by an accretion disk at 6 au from the
central object. Such a  brown dwarf+disk system could have the Keplerian velocity and extension needed to explain the moving
shadow and would still escape a detection. Additional Jovian planets at larger radii could account for the observed properties of
J1604 while explaining our non-detection. The expected details of our proposed multiple system configuration are yet to be explored
by models.

Our observations reveal, for the first time, the disk around J1604 in scattered light at all $Y, J,H, K$ bands. Despite
being affected by artefacts, we show that the disk color is red, which is best explained by dust particles with an
average size $\gtrsim0.3 \mu$m. We find that the brightness radial profile changes its slope at $\sim79$ au, which
matches the cavity size radius for the mm-sized particles derived by \citet{2014ApJ...791...42Z}. This suggests
that there is a relationship between the change in the flaring angle of the disk (traced by the scattered light images
sensitive to the small dust grains in the disk surface), and the radial distribution of the large grains in the disk
midplane. 

Overall, J1604 shows very different properties than the majority of well-known transitional (and pre-transitional) disks.
Its extremely low accretion rate \citep[$\dot{M}_{acc} <10^{-11} \MSun$ yr$^{-1}$,][]{2012ApJ...753...59M} seems to be in
contradiction with the large amount of gas mass \citep[$M_{gas}\sim 0.02 \MSun$, ][]{2015A&amp;A...579A.106V}
still present in this 5-12 Myr old disk. The gas cavity is the most gas-depleted cavity in the sample studied by
\citet{2015A&amp;A...579A.106V} and \citet{2016A&amp;A...585A..58V}. Finally we note that, although multiple
planets are currently the most realistic explanation, photoevaporation and other mechanisms (e.g., a dead-zone)
could also be creating the observed features.

%-------------------------------------------------------------------
\begin{acknowledgements}
        We thank the referee for his/her useful comments.
        H.C., E.V., and G.M. acknowledges support from the Spanish Ministerio de Econom\'ia y Competitividad under grant AYA 2014-55840-P.
        GM is funded by the Spanish grant RyC-2011-07920. This research was partially funded by the Millennium Science Initiative, Chilean
        Ministry of Economy, Nucleus RC130007. CC acknowledges support from CONICYT PAI/Concurso nacional de inserci\'on en la academia,
        convocatoria 2015, Folio 79150049, and CONICYT-FONDECYT grant 3140592.
        MRS acknowledge support from FONDECYT grant 1141269.  LC was supported by ALMA-CONICYT and CONICYT-FONDECYT
        31120009 and 1140109.
\end{acknowledgements}

\bibliographystyle{aa.bst}      % (uses file "plain.bst")
\bibliography{j1604.bib}         % expects file "myrefs.bib"

%\clearpage
%\newpage

%\Online
%\begin{appendix} %First online appendix
%\end{appendix}

\end{document}